\documentclass[apj]{emulateapj}
\usepackage{psfig}

\newcommand{\etal}{et al.}

\newcommand\asca{{\it ASCA\/}}
\newcommand\chandra{{\it Chandra}}
\newcommand\xmm{{\it XMM-Newton\/}}

\newcommand\suzaku{{\it Suzaku\/}}
\newcommand\cco{CXOU~J171405.7$-$381031}

\newcommand\ctb{CTB~37B}
\newcommand\tevtwo{HESS~J1713$-$381}
\newcommand\rxj{RXJ 1713.7$-$3946}
\newcommand\snr{G353.6$-$0.7}
\newcommand\tev{HESS~J1731$-$347}
\newcommand\psr{XMMU~J173203.3$-$344518}

\def\simlt{\mathrel{\hbox{\rlap{\hbox{\lower4pt\hbox{$\sim$}}}\hbox{$<$}}}}
\def\simgt{\mathrel{\hbox{\rlap{\hbox{\lower4pt\hbox{$\sim$}}}\hbox{$>$}}}}

\slugcomment{}

\shorttitle{Magnetar in SNR CTB 37B}
\shortauthors{Halpern \& Gotthelf}

\begin{document}

\title{An Energetic Magnetar in HESS J1713$-$381/CTB 37B}

\author{J. P. Halpern and E. V. Gotthelf}
\affil{Columbia Astrophysics Laboratory, Columbia University,
550 West 120th Street, New York, NY 10027-6601, USA;
jules@astro.columbia.edu, eric@astro.columbia.edu}

\begin{abstract}
We obtained a second \chandra\ timing measurement of the 3.82~s pulsar
\cco\ in the supernova remnant (SNR) \ctb, which shows that it is
spinning down rapidly.  The average period derivative of
$(5.88 \pm 0.08) \times 10^{-11}$ over the 1 year time span corresponds
to a dipole magnetic field strength $B_s = 4.8 \times 10^{14}$~G,
well into the magnetar range.  The spin-down power
$\dot E = 4.2 \times 10^{34}$ erg~s$^{-1}$
is among the largest for magnetars, and the 
corresponding characteristic age
$\tau_c \equiv P/2\dot P = 1030$~years is comparable to estimates
of the age of the SNR.  The period derivative enables
us to recover probable pulsations in an \asca\ observation taken in
1996, which yields a mean characteristic age of 860 years over the
longer 13 year time span.
The source is well detected up to 10~keV,
and its composite spectrum is typical of a magnetar.
\ctb\ hosts \tevtwo,
the first TeV source that is coincident with a magnetar.
While the TeV emission has been attributed to the SNR shell,
it is possibly centrally peaked, and we hypothesize that this particularly
young, energetic magnetar may contribute to the HESS source.
We also searched for pulsations from another source in a HESS SNR,
\psr\ in \tev/\snr, but could not confirm pulsations or 
long-term flux variability, making it more likely that this source
is a weakly magnetized central compact object.

\end{abstract}


\section {Introduction}

\begin{figure}[b]
\centerline{\psfig{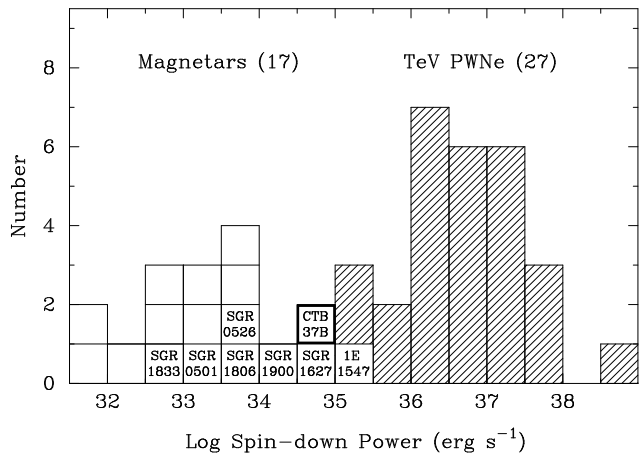}}
\vspace{0.1in}
\caption{
Spin-down power of 27 pulsars (hatched)
whose PWNe are identified as TeV sources
by HEGRA, HESS, VERITAS, MAGIC, and Milagro, in comparison with spin-down
power of 17 magnetars (open squares)
with measured period derivatives.  The TeV PWNe
are drawn from the reviews referenced in Section 1, and
magnetars from references in Footnote 2.  Magnetars that have had SGR
outbursts, including 1E~1547.0$-$5408, are labeled individually;
the unlabeled squares are AXPs.
The spin-down power of \cco\ in \ctb\ is among the largest of magnetars,
and falls in the range occupied exclusively by SGRs, although it is
not known to be an SGR.
}
\label{hist}
\end{figure}

Of the $\approx 60$ Galactic TeV
sources\footnote{VHE $\gamma$-ray Sky Map and Source Catalog,
http://www.mppmu.mpg.de/\~\,rwagner/sources/index.html},
the majority (43) are identified as either pulsar wind nebulae (PWNe)
or supernova remnants (SNRs).  The PWN class comprises
$\approx 27-30$ of these, and it is likely that some of the
unclassified sources will also prove to be PWNe.
Most pulsars that are responsible for the
PWNe detected at X-ray and TeV energies have spin-down power
$\dot E \equiv 4\pi^2I\dot P/P^3 > 10^{36}$ erg~s$^{-1}$.
In Figure~\ref{hist} we display the distribution of $\dot E\/$
for rotation-powered pulsars powering TeV nebulae.
For reviews of these, see \citet{car08}, \citet{gal08},
\citet{lem08}, \citet{hes08}, \citet{mat09}, \& \citet{kar10}.
In the standard model of nonthermal emission from PWNe,
the X-rays are synchrotron
from relativistic electrons and positrons, while 
the TeV $\gamma$-rays are inverse Compton scattered
microwave background and other ambient photons
from the same population of relativistic particles
\citep[e.g.,][]{dej08,zha08}.
In contrast, there is scant evidence that any of the 17
known magnetars\footnote{SGR/AXP Online Catalog,
http://www.physics.mcgill.ca/\~\,pulsar/magnetar/main.html},
with periods in the range 2$-$12~s, produce X-ray or TeV
PWNe.  Including both anomalous X-ray pulsars (AXPs)
and soft gamma-ray repeaters (SGRs), the spin-down power
of magnetars is $< 10^{36}$ erg~s$^{-1}$, 
as also plotted in Figure~\ref{hist}.
Because of their larger magnetic field strengths and different
emission mechanisms from ordinary spin-powered pulsars, it is not
predicted that magnetars accelerate particles to TeV energies
\citep{bel07,tho08a,tho08b}, except perhaps in their earliest stage
of rapid spin-down \citep{aro03}.

\citet[][hereafter Paper I]{hal10a} presented a study of
point X-ray sources in two SNRs detected at TeV energies
by the HESS array of atmospheric Cherenkov telescopes.
The first of these, \tevtwo\ associated with \ctb,
was discovered by \citet{aha06}.  Using a \chandra\
observation, \citet{aha08a} then found the point X-ray source
\cco\ in \ctb, and considered that it could be a pulsar,
albeit with an unusually soft, non-thermal spectrum.
\citet{nak09} analyzed \chandra\ and \suzaku\ spectra of
\cco, noting possible variability in flux,
which they took to be good evidence that it is an
anomalous X-ray pulsar (AXP).  In Paper~I, we
reported the discovery of pulsations from \cco\ that
verifies this conjecture.
In Section~2 of this paper,
we report a follow-up \chandra\ observation that
measures the period derivative of the pulsar, establishing
its quantitative magnetar properties.\footnote
{While this paper
was in preparation, the same result was announced
by \citet{sat10} using an \xmm\ observation.}
We raise in
Section~3 the possibility that at an early stage,
a magnetar such as the one in \ctb\ may produce a TeV PWN.
In Section~4 we report on a new observation of the second
compact X-ray source discussed in Paper~1, 
\psr\ in \tev/\snr, in which pulsations are not confirmed.

\begin{deluxetable*}{lcccccl}
\tablewidth{0pt}
\tablecaption{Log of Timing Observations of \cco\ in \ctb}
\tablehead{
\colhead{Instrument/Mode} & \colhead{ObsID} & \colhead{Date (UT)}
& \colhead{Date (MJD)} & \colhead{Exp. (s)} & \colhead{Counts (s$^{-1}$)}
& Period (s)
}
\startdata
\asca\ GIS          & 54002030 & 1996 Sep 12 & 50,338.8 & 13,322 & 0.053  & $3.7954(1)$ \\
\chandra\ ACIS-S3/CC/F  & 10113 & 2009 Jan 25 & 54,856.3 & 30,146 & 0.106 & $3.823056(18)$ \\
\chandra\ ACIS-S3/CC/F  & 11233 & 2010 Jan 30 & 55,226.5 & 30,124 & 0.122 & $3.824936(17)$ \\
\xmm\ EPIC pn           & 0606020101 & 2010 Mar 17 & 55,272.2 & 40,264 & 0.264 & $3.825352(4)\tablenotemark{a}$ 
\enddata
\label{logtable}
\tablenotetext{a}{From Sato et al. (2010).}
\end{deluxetable*}

\begin{figure}[b]
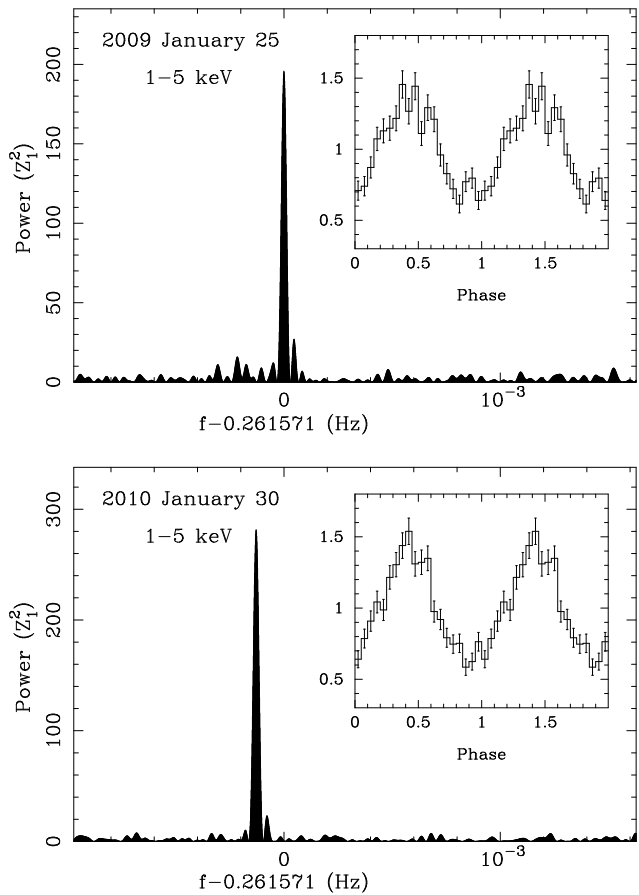

\vspace{0.1in}
\centerline{\psfig{figure=f2a.eps,width=0.97\linewidth,angle=270}}
\vspace{0.15in}
\centerline{\psfig{figure=f2b.eps,width=0.97\linewidth,angle=270}}
\caption{
$Z_1^2$ power spectra of \cco\ from the \chandra\ CC-mode
observations separated by 1 year, and corresponding folded light curves
(inset).  Normalized count rate is plotted.  Phase is shifted
arbitrarily. Note the highly significant decrease in frequency.
}
\label{pulse}
\end{figure}

\section {X-ray Observations of \cco\ in \tevtwo/\ctb}

We obtained a new \chandra\ observation of \cco\ on 2010 January 30
using the same instrumental setup as our previous observation of
2009 January 25 (Paper~I).  Both data sets were acquired using
the Advanced Camera for Imaging and Spectroscopy (ACIS) in
continuous-clocking (CC) mode with the pulsar located on the ACIS-S3
CCD.  This provides a time resolution of 2.85~ms and avoids any
spectral pile-up.  The photon arrival times in CC mode are adjusted
in the standard processing to account for the known position of the pulsar,
spacecraft dither, and SIM offset. Reduction and analysis used the standard
software packages CIAO (v4.1) and CALDB (v3.5.0).
A log of the \chandra\ timing observations is given in Table~\ref{logtable}.

With a period derivative derived from the \chandra\ observations
(see below), we also searched for and likely detected pulsations in archival
data taken with the \asca\ X-ray telescope \citep{tan94}.
The pulsar was imaged with the Gas Imaging Spectrometer (GIS) on
1996 September 12 as part of the Galactic Survey \citep{sug01}.
It fell on the very edge of the GIS field of view of sequence \#54002030,
$21'$ from the pointing direction, partially in the area masked out
by the standard processing. To maximize the sensitivity to the expected
pulsar signal, we combined unmasked and unfiltered data taken in low-
and high-bit-rate data modes, with timing resolution of 0.5~s and
0.0625~s, respectively. We obtained a total of 712 counts (0.5$-$10~keV)
from the source in a $3'$ radius aperture during $13.3$~ks of usable
exposure from the 21~ks observation.

\subsection{Timing Analysis}

The photon arrival times were corrected to the solar system
barycenter using the pulsar position obtained from the
\chandra\ ACIS-I image.
Timing analysis of the \chandra\ data
was then performed in an identical manner
to Paper~I.  Photons from the central four source columns
($2^{\prime\prime}$) in the range 1$-$5~keV,
which contains most of the photons,
were extracted and searched for pulsations using the
Rayleigh test \citep{str80}, also known as $Z_1^2$ \citep{buc83}.
The power-spectrum and folded pulse profiles from the two \chandra\ 
timing observations are shown in Figure~\ref{pulse}.
Although the eye may be drawn to subtle
differences, a $\chi^2$ test on the difference profile
indicates that the two light curves are not significantly different:
$\chi^2_{\nu} = 1.44$ for 20 degrees of freedom, corresponding
to 9\% chance probability.
The pulsed fraction is $f_p \approx 0.38$, defined as the fraction
of counts above the minimum in the light curve, corrected for background.
There is no evidence for energy dependence of the pulse profile.
While it is possible that the true period is 7.6~s,
with two peaks per rotation, there is no strong evidence to
support this in the form of distinguishable peaks in the current
data.

\begin{deluxetable}{ll}[t]
\tablewidth{0pt}
\tablecaption{\chandra\ Measured Spin Parameters of \cco\ in \ctb}
\tablehead{
\colhead{Parameter} & \colhead{Value}
}
\startdata
R.A. (J2000)\tablenotemark{a}         & $17^{\rm h}14^{\rm m}05^{\rm s}\!.74$ \\
Decl. (J2000)\tablenotemark{a}        & $-38^{\circ}10^{\prime}30^{\prime\prime}\!.9$ \\
Epoch (MJD)                           & 55226.5\\
Spin period, $P$                      & 3.824936(17)~s \\
Period derivative, $\dot P$           & $(5.88 \pm 0.08) \times 10^{-11}$ \\
Range of dates (MJD)                  & 54856$-$55227 \\
Surface dipole magnetic field, $B_s$  & $4.8 \times 10^{14}$~G \\
Spin-down power, $\dot E$        & $4.2 \times 10^{34}$~erg~s$^{-1}$ \\
Characteristic age, $\tau_c$          & 1030~yr
\enddata
\tablenotetext{a}{Measured from \chandra\ ACIS-I ObsID 6692.
Typical \chandra\ ACIS coordinate uncertainty is $0\farcs6$.}
\label{ephemeris}
\end{deluxetable}

\begin{figure}[t]
\centerline{\psfig{figure=f3a.eps,width=0.97\linewidth,angle=0}}
\vspace{0.2in}
\centerline{\psfig{figure=f3b.eps,width=0.97\linewidth,angle=0}}
\caption{
(Top) Period measurements of \cco\ from the two \chandra\ CC-mode observations
(filled circles) and the one \xmm\ observation 
(Stato \etal\ 2010, open circle).  Error bars are smaller than the size
of the symbols.  Period derivatives between consecutive measurements
are indicated.  A significant increase in $\dot P$ is seen.
(Bottom) Average period derivative from 1996 to 2010 when
including the probable \asca\ GIS detection in 1996.  The box
encloses the range of the top panel.
}
\vspace{0.1in}
\label{timing}
\end{figure}

\begin{figure}[t]
\centerline{\psfig{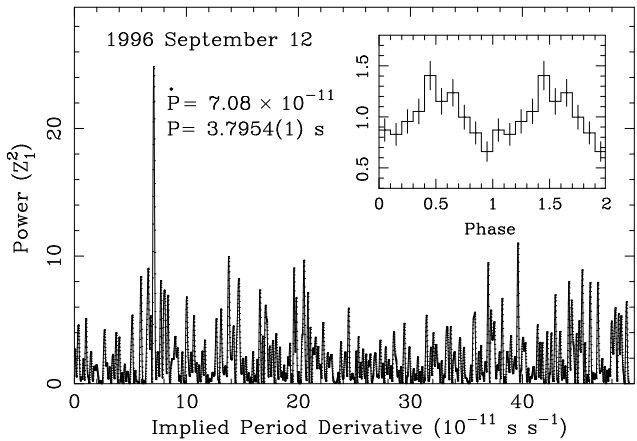}}
\caption{
Period search of \cco\ in the \asca\ GIS observation of 1996 September 12.
Period is converted to the $\dot P$ needed to extrapolate from the
\chandra\ observation of 2009. Inset: Folded light curve in the
total (0.5$-$10~keV) energy band corresponding to the indicated
highest peak in the power spectrum.  Normalized count rate is plotted.
}
\label{asca}
\end{figure}

The periods and $1\sigma$ uncertainties for the two \chandra\
observations are listed in Table~\ref{logtable}, while the
period derivative listed in Table~\ref{ephemeris} is
calculated from the difference between the periods measured
1 year apart.  The uncertainty on $\dot P$ is propagated from the 
statistical errors in the individual period measurements.
The $\dot P = (5.88 \pm 0.08) \times 10^{-11}$ over the 1 year
time span corresponds to a dipole magnetic field strength
$B_s = 3.2 \times 10^{19}(P\dot P)^{1/2} = 4.8 \times 10^{14}$~G,
well into the magnetar range.  The spin-down power
$\dot E = 4.2 \times 10^{34}$ erg~s$^{-1}$
is among the largest for magnetars,
and the characteristic age $\tau_c \equiv P/2\dot P = 1030$~years
is among the smallest.

A third period measurement was obtained by \citet{sat10} using
\xmm, only 47 days after our second \chandra\ timing observation.
They reported $P = 3.825352 \pm 0.000004$~s on 2010 March 17$-$18.
As shown in Figure~\ref{timing}, this third point
requires a significant increase in $\dot P$ to
$1.05 \times 10^{-10}$.  Such variation is common in
magnetars, and in this case may indicate that the spin-down
power was recently as large as $\approx 7 \times 10^{34}$ erg~s$^{-1}$.

We recovered a fourth period measurement from the archival \asca\
observation listed in Table~\ref{logtable} by extrapolating
the \chandra\ ephemeris.   This detection shown in
Figure~\ref{asca}, while not significant
in a blind search, has a false detection probability of
only $1.1 \times 10^{-3}$ when 
the search is restricted to a range of $\dot P$ up to
twice the \chandra\ measured value.  The implied period
derivative relative to the 2009 \chandra\ measurement is
$\dot P = 7.08 \times 10^{-11}$, and the strength
of the highly modulated signal is plausible.
As shown in Figure~\ref{asca},
which includes an extended range of period derivatives
up to 10 times the \chandra\ value,
no other significant peak is found.
Figure~\ref{timing} shows the average $\dot P$ obtained in
an unweighted fit to all four exsiting
period measurements, $\dot P = 7.03 \times 10^{-10}$
corresponding to $\tau_c = 860$~years,
very close to the \chandra\ measured values, but perhaps
a better representation of the long-term average.
Keeping in mind that $\dot P$ is evidently variable,
the ephemeris in Table~\ref{ephemeris} includes
only the two \chandra\ points,
which present the minimum observed value of $\dot P$.

\subsection{Spectral Analysis}

Spectral data for the pulsar were extracted from the \chandra\
CC-mode observations using the
five central source columns, corresponding to a diameter
of $2.\!^{\prime\prime}46$ and $\approx 95$\% of the
point-source energy enclosed.
The background for the spectrum
was obtained from the regions adjacent to the source. 
Background subtracted count rates are
$0.106 \pm 0.002$ and $0.122 \pm 0.002$ s$^{-1}$, respectively,
for the two observations, indicating a likely change of $\approx 15\%$.

\begin{figure*}[]
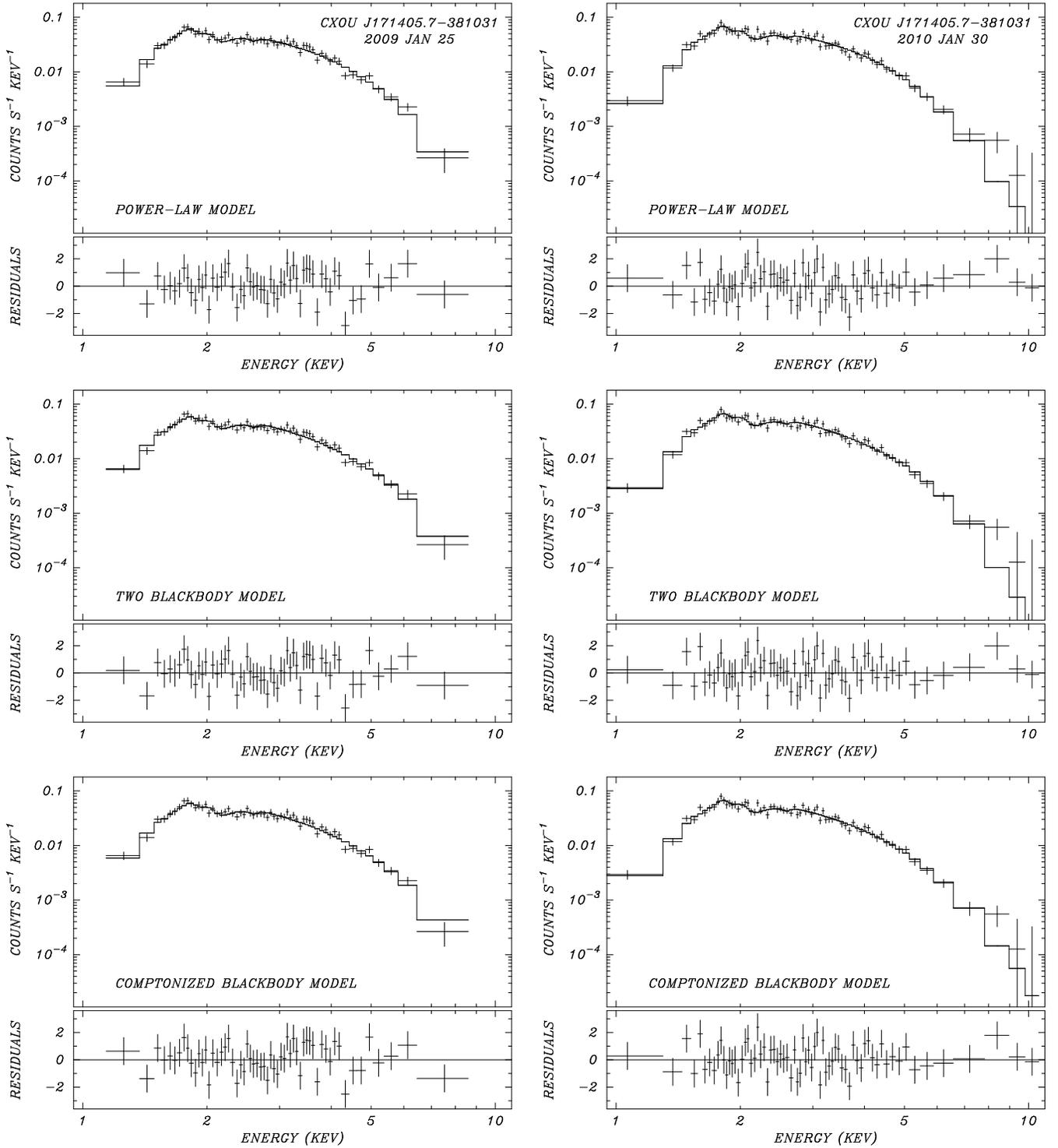

\centerline{\psfig{figure=f5a.eps,width=0.48\linewidth,angle=270}
\hspace{0.1in}
\psfig{figure=f5b.eps,width=0.48\linewidth,angle=270}}
\vspace{0.1in}
\centerline{\psfig{figure=f5c.eps,width=0.48\linewidth,angle=270}
\hspace{0.1in}
\psfig{figure=f5d.eps,width=0.48\linewidth,angle=270}}
\vspace{0.1in}
\centerline{\psfig{figure=f5e.eps,width=0.48\linewidth,angle=270}
\hspace{0.1in}
\psfig{figure=f5f.eps,width=0.48\linewidth,angle=270}}
\caption{
Spectra of \cco\ in \ctb\ from the \chandra\ ACIS-S3 CC-mode
observations of 2009 (left) and 2010 (right) 
fitted to power-law (top), two blackbody (middle),
and Comptonized blackbody models (bottom).
Fitted parameters are listed in Table~\ref{spectable}.
}
\label{spectrum}
\end{figure*}

The standard
spectral response matrices were generated for the target location on
the CCD using the CIAO tool {\tt psextract}, with care taken
to normalize the background according to area.  All spectra were corrected
for the effects of charge-transfer inefficiency; however, the spectral
gain is not calibrated well in CC mode. The resulting spectra were
grouped with a minimum of 50 counts per bin and fitted using the
XSPEC software \citet{arn96} to power-law, two blackbody,
and Comptonized blackbody models.  The spectral fits are shown
in Figure~\ref{spectrum}, and the best fitted
parameters for these models are listed in Table~\ref{spectable}.
A single blackbody model does not yield a good fit, and is not shown.

In paper~I the ACIS-I/TE mode observation of 2007
indicated a rising spectrum above 6~keV, but the CC-mode spectra
presented here do not clearly show that property.  We suspect that the
previous spectrum may have suffered systematic effects
as the pulsar was dithered into the gap between CCDs.
In addition, a small amount of spectral pileup in TE mode is
avoided in CC mode.
Nevertheless, the source is definitely detected up to 10~keV,
and shows some evidence for an upturn there in the slightly 
brighter state of 2010.

\begin{deluxetable}{lcc}[t]
\tablewidth{0pt}
\tablecaption{\chandra\ Spectral Fits for \cco\ in \ctb}
\tablehead{
\colhead{Parameter} & \colhead{2009 Jan 25} & \colhead{2010 Jan 30}
}
\startdata
& Power-Law Model\\
\tableline
$N_{\rm H}$~($10^{22}$ cm$^{-2}$)         & $5.46 \pm 0.36$ & $5.36 \pm 0.34$ \\
$\Gamma$                                  & $4.25 \pm 0.23$ & $4.12 \pm 0.22$ \\
$F_x(2-10\ {\rm keV}$)\tablenotemark{a} & $1.26 \times 10^{-12}$  &  $1.48 \times 10^{-12}$ \\
$L_x(2-10\ {\rm keV}$)\tablenotemark{b} & $2.4 \times 10^{34}$   & $2.7 \times 10^{34}$ \\
$\chi^2_{\nu}(\nu)$                       & 1.06(56)               & 1.07(67) \\
\tableline
& Two Blackbody Model \\
\tableline
$N_{\rm H}$~($10^{22}$ cm$^{-2}$)         & $3.81^{+0.44}_{-0.27}$ & $4.01^{+0.67}_{-0.28}$ \\
$kT_1$ (keV)                              & $0.45^{+0.02}_{-0.08}$ & $0.43^{+0.03}_{-0.08}$ \\
$R_1$ (km)                                & $2.6_{-0.9}^{+3.1}$    & $3.2_{-1.2}^{+3.9}$ \\
$kT_2$ (keV)                              & $1.13^{+0.48}_{-0.27}$ & $1.13^{+0.38}_{-0.23}$ \\
$R_2$ (km)                                & $0.19_{-0.13}^{+0.32}$ & $0.22_{-0.14}^{+0.28}$ \\
$F_x(2-10\ {\rm keV}$)\tablenotemark{a} & $1.28 \times 10^{-12}$  &  $1.50 \times 10^{-12}$ \\
$L_x(2-10\ {\rm keV}$)\tablenotemark{b} & $1.8 \times 10^{34}$   &  $2.2 \times 10^{34}$ \\
$L_{\rm bol}$\tablenotemark{b} & $4.2 \times 10^{34}$   &  $5.4 \times 10^{34}$ \\
$\chi^2_{\nu}(\nu)$                       & 1.11(54)               & 1.07(65) \\
\tableline
& Comptonized Blackbody Model\\
\tableline
$N_{\rm H}$~($10^{22}$ cm$^{-2}$)         & $4.12^{+0.50}_{-0.41}$ &  $4.11^{+0.82}_{-0.34}$ \\
$kT$ (keV)                              & $0.38^{+0.08}_{-0.05}$   &  $0.38^{+0.08}_{-0.05}$ \\
$R$ (km)                                & $3.3_{-1.6}^{+3.4}$      &  $3.5_{-1.3}^{+9.0}$  \\
$\Gamma$                                  & $3.34^{+0.18}_{-0.20}$   & $3.22^{+0.32}_{-0.15}$ \\
$F_x(2-10\ {\rm keV}$)\tablenotemark{a} & $1.32 \times 10^{-12}$  &  $1.57 \times 10^{-12}$ \\
$L_x(2-10\ {\rm keV}$)\tablenotemark{b} & $ 1.8 \times 10^{34}$   &  $2.2 \times 10^{34}$ \\
$\chi^2_{\nu}(\nu)$                       & 1.10(55)               &  1.04(66)
\enddata
\tablenotetext{a}{Absorbed flux in units of erg cm$^{-2}$ s$^{-1}$.}
\tablenotetext{b}{Unabsorbed luminosity in units of erg s$^{-1}$,
assuming $d = 8$~kpc.}
\label{spectable}
\end{deluxetable}

\section{Discussion}

\subsection{\cco\ as a Magnetar}

The \chandra\ CC-mode spectra of \cco\ are fitted equally well by
power-law, two blackbody, and Comptonized blackbody
models.  Of these, we disfavor the steep $\Gamma \approx 4.2$
power-law model as being least plausible on physical grounds.
In the thermal models, there is a blackbody component
of temperature $\approx 0.4$~keV that covers $\approx 10\%$
of the surface area of the neutron star.  A harder component
is also needed that can be parameterized as either a second
blackbody or a power-law tail, the latter approximating
what is probably cyclotron resonance scattering \citep{fer07}
using a simplified analytic expression in the Comptonized
blackbody model \citep[][and references therein]{hal08}.
The hot spot(s), and
possibly anisotropic scattering in the magnetosphere, 
are responsible for the observed pulsations, although
there is no obvious dependence of the pulse shape on energy. 
In either the two blackbody
or the Comptonized blackbody model, the associated
$N_{\rm H} \approx 4 \times 10^{22}$~cm$^{-2}$ is compatible with
its value in fits to the thermal emission from the SNR \citep{nak09},
which yield $N_{\rm H} =3.5^{+0.5}_{-0.7}\times 10^{22}$~cm$^{-2}$.
This provides some additional support for the association
of \cco\ with \ctb.

Other properties of \cco\ are consistent with those of AXPs:
a factor of $\approx 1.8$ flux decrease is reported between
the earlier \suzaku\ and \chandra\ observations \citep{nak09}.
The spin-down power of \cco,
$\dot E \geq 4.2 \times 10^{34}$ erg~s$^{-1}$, is among the
highest for magnetars, exceeded only by SGR/AXP 1E~1547.0$-$5408
\citep[$\dot E \geq 1 \times 10^{35}$ erg~s$^{-1}$;][]{cam07,cam08},
and equalled by SGR~1627$-$41
\citep[$\dot E = 4.3 \times 10^{34}$ erg~s$^{-1}$;][]{esp09}
and SGR~1806$-$20 when in its active state,
$\dot E \sim 5 \times 10^{34}$ erg~s$^{-1}$,
although here we adopt a more typical quiescent value of
$8 \times 10^{33}$ erg~s$^{-1}$ \citep{woo07}.
The range of $\dot E\/$ in which \cco\ falls
(Figure~\ref{hist}) is occupied by SGRs rather than AXPs.
By analogy with the history of 1E~1547.0$-$5408 and other
high-$\dot E\/$ magnetars that exhibit large fluctuations in $\dot P$,
we speculate that \cco\ is prone to having SGR outbursts.
It also shares with 1E~1547.0$-$5408 the property 
that its spin-down power is greater than its X-ray luminosity
($L_x \approx 2 \times 10^{34}$ erg~s$^{-1}$),
at least in quiescence.  However, this does not change our assessment
that the majority of its luminosity is generated by magnetic
field decay and not by rotation, given its high surface dipole
field strength, $B_s = 4.8 \times 10^{14}$~G, X-ray spectral
properties, and variability that are typical of magnetars.

\subsection{Comparison of Age Estimates}

At radio wavelengths, \ctb\ is a shell $\approx 10^{\prime}$ in diameter
\citep{kas91}.  Its distance, estimated from \ion{H}{1} absorption
and the Galactic rotation curve, is $\approx 8$~kpc \citep{gre09}.
Estimates of the age of \ctb\ are quite varied, and it is of great
interest to compare these with the spin-down age of \cco,
even though characteristic ages of magnetars are not reliable
because their period derivatives are known to vary by a factor
of a few.

\citet{cla75} tentatively associated \ctb\ with the
historical supernova of 393~AD, which would make it 
$1600$ years old, almost twice the pulsar's characteristic
age of $\sim 860$ years.  However, in the same region of the sky
the more recently discovered SNR
\rxj\ is a much better candidate for SN~393,
as argued by \citet{wan97}, \citet{wan06}, and \citet{cas04}.
At $d \approx 1.3$~kpc and with $N_{\rm H}
\approx 5 \times 10^{21}$~cm$^{-2}$
\citep{cas04}, \rxj\ is closer, and an order of magnitude less
absorbed than \ctb, while the latter is too far away to
have been visible for 8 months as the Chinese records indicate.
Given this better alternative,
we do not regard the birth of \cco\ as having
been established by historical records.  Instead, we turn to
estimates of the SNR age from analyses of its thermal X-ray spectrum.

Using the \chandra\ data, \citet{aha08a} fitted a non-equilibrium
ionization model and obtained a Sedov age of $\approx 4900$~yr
for \ctb\ assuming electron-ion equilibrium, while noting that
this condition is usually not obtained in young SNRs.
Both a higher ion temperature and efficient
cosmic ray acceleration have the effect of decreasing the
inferred age, perhaps to $\sim 2700$~yr in the estimate of
\citet{aha08a}.  \citet{nak09} then analyzed the \suzaku\ spectrum
of \ctb\ and estimated the age of the plasma
as $650^{+2500}_{-300}$~years in the brightest part of the remnant,
which also required a relatively low ISM density of
$n_{\rm H} \sim 0.4$~cm$^{-3}$.
In addition, they found a region of non-thermal power-law emission
and a smaller ambient density in the southern part of the SNR.
Thus, the SNR age estimate is broadly consistent with the pulsar's
characteristic age, although the uncertainties in both values are
large, and their applicability can be questioned.

\subsection{Magnetar/SNR Associations}

Three magnetars are securely associated
with shell SNRs because of their central locations:
1E~2259+586 in CTB~109, 1E~1841$-$045 in Kes~73,
SGR 0526$-$66 in the LMC SNR N49 (Gaensler \etal\ 2001).
To these we may tentatively add
1E~1547.0$-$5408 because of its possible radio shell G327.24$-$0.13
\citep{gel07}, and now \cco\ in \ctb.
In addition to constraining ages
for modelling of PWNe,
such associations are valuable for inferring
natal properties such as magnetar kick velocity,
and possible contribution of an assumed
rapidly spinning proto-magnetar
to the energetics of the SNR expansion (e.g., Vink \& Kuiper 2006).
Theories for the birth of magnetars require short initial
spin periods, of order a few milliseconds, in order to
create their strong $B$-fields from a
turbulent dynamo whose strength depends on the rotation
rate of the proto-neutron star \citep{tho93}.

The radio emitting AXP XTE~J1810$-$197 is the only
magnetar with a directly measured proper motion,
corresponding to $v_t = 212 \pm 35\ d_{3.5}$ km~s$^{-1}$ \citep{hel07},
a value which is not unusual compared to ordinary young neutron stars.
\citet{aha08a} noted that the
displacement of \cco\ from the center of the shell
of \ctb\ would require a transverse velocity of
$\sim 1000$ km~s$^{-1}$ for an age of 5000~yr.
The velocity would be unreasonably larger if the 
pulsar characteristic age of $\sim 1000$~years is adopted
as its true age.  However, \citet{aha08a} and \citet{nak09}
also noted that the brighter X-ray and radio
emission on the eastern side of the remnant
(see Figure~1 of Paper~I) is probably due to a higher
density there, implying that the pulsar may still be close
to the explosion center, which is not the geometric center,
reducing its inferred velocity.

Because of this non-uniformity
in shock radius $r_s$ and ISM density $\rho_0$,
it is difficult to use the standard Sedov evolution
$$r_s^5(t) = {2.026\,\mathcal{E}_0\,t^2 \over \rho_0}$$
as did \citet{vin06} for other SNRs hosting magnetars,
to estimate the energy $\mathcal{E}_0$ of the explosion.
The radius, which enters to the fifth power, is uncertain
by at least a factor of 2.  Nevertheless,
if we use the values adopted by \citet{nak09}
for the radius, age, and ambient density in the
brightest part of the remnant (their region 1),
an unremarkable energy of $\mathcal{E}_0 \sim 1 \times 10^{50}$~erg results.
Energetic input to the SNR expansion
from a presumed millisecond magnetar spinning down
is therefore not evident, as was also concluded by \citep{vin06}
for other SNRs hosting magnetars.

\subsection{Can the Magnetar Contribute to TeV Emission?}

HESS observations reported in \citet{aha08a} localized the
TeV emission to the center of the radio shell of \ctb, and
found that the extent of the TeV source is compatible with
either a centrally peaked Gaussian of 
$\sigma = 2.\!^{\prime}6 \pm 0.\!^{\prime}8$
or a shell of radius $4^{\prime}$--$6^{\prime}$,
consistent with the size of the radio remnant.
The energy flux of \tevtwo\ from 0.2--10~TeV
is $\approx 4.2 \times 10^{-12}$ erg~cm$^{-2}$~s$^{-1}$,
corresponding to luminosity
$L_{\gamma} = 3.2 \times 10^{34}\ d_8^2$ erg~s$^{-1}$,
which we note is almost equal to the spin-down power
of \cco, $\dot E = 4.2 \times 10^{34}$ erg~s$^{-1}$.
Previous authors have argued that particles accelerated at the
SNR shock are responsible for the TeV emission from \tevtwo\
via pion decay \citep{aha08a} or multi-zone inverse Compton
scattering of the microwave background \citep{nak09}.
As it is not clear if the
TeV source has a pure shell-like morphology,
it cannot be ruled out that the pulsar
makes some contribution to the TeV flux,
possible via inverse Compton scattering from a PWN
that is confined within the SNR shell.

Since the TeV luminosity of \tevtwo\ is almost equal to 
the {\it present\/} spin-down power of \cco, a scenario
for a pulsar-powered TeV nebula would necessarily involve
particles injected at an earlier time when its spin-down
power was larger. However, unlike ordinary pulsars, it is
not certain that magnetars can accelerate 
particles to TeV energies.  Magnetar models that involve strong currents
on closed, twisted magnetic field lines develop voltages of only
$\sim 10^9$~V \citep{bel07}.  On the other hand, it is not excluded
that the ordinary pulsar mechanism operates on open magnetic field lines
of magnetars, which may generate particle-dominated winds that
become shocked PWNe, accelerating a fraction of the particles to
TeV energies.
Evidence for this possibility comes from the three magnetars
that are also transient radio pulsars \citep{cam06,cam07,lev10}.
Even so, we caution that models for radio emission from magnetars
\citep{tho08a,tho08b} produce average particle energies of
only $\sim 3$~GeV even on open field lines, 
because $\gamma$-rays make pairs more easily
in the strong magnetic field.

Nevertheless, if high-energy electrons are generated
in a shocked PWN in an early stage
and escape the high $B$-field region,
they can emit via inverse Compton
scattering into the TeV band for an extended time
\citep{dej08}. Electron lifetimes against synchrotron and
inverse Compton losses are
$$t_s \approx {4200\ {\rm yr} \over \sqrt{E_{\gamma}({\rm TeV})}}
\left({B \over 10\ \mu{\rm G}}\right)^{-2}$$
and
$$t_{\rm IC} \approx {8 \times 10^4\ {\rm yr}
\over \sqrt{E_{\gamma}({\rm TeV})}},$$
respectively, for an electron producing a photon
of energy $E_{\gamma}({\rm TeV})$
by inverse Compton scattering of the microwave background.  The
limiting lifetime of such a relic nebula is then determined by $t_s$
as long as $B>2\ \mu$G.

The spin-down power of \cco\ was certainly large enough to generate
a TeV PWN during the history of \ctb, as long as it was born with a
period significantly shorter than its present value, since
$\dot E = 9 \times 10^{36}\,P^{-4}\,(B_s/5 \times 10^{14}~{\rm G})^2$
erg~s$^{-1}$.  The current TeV
luminosity of \tevtwo, $\approx 3.2 \times 10^{34}$ erg~s$^{-1}$,
if radiated for 1000~yr, amounts to $\sim 1 \times 10^{45}$~erg,
which is $< 10^{-3}$ of the initial rotational energy of
the pulsar for $P_0 < 0.14$~s.
Thus, we consider an inverse Compton
PWN to be a plausible channel for TeV emission from \cco\
because it requires only a small fraction of the rotational
energy to be deposited in very high-energy electrons.

A possible prototype of such systems is PSR~J1846$-$0258, the
``transitional'' $\sim 700$~yr old pulsar \citep{got00}
associated with HESS~J1846$-$029 in the shell-like SNR Kes~75
\citep{dja08}. This $0.32$~s pulsar has a dipole field that
approaches magnetar strength ($B_s = 4.9 \times 10^{13}$~G) and
displays AXP-like bursts (Gavriil et al. 2008); it is therefore likely
of an intermediate class connecting the rotation-powered and the
magnetar pulsars.  It is also one of the most energetic pulsars;
its $\dot E =  8.1 \times 10^{36}$ erg~s$^{-1}$ is
easily sufficient to power its X-ray PWN and the
compact TeV emission of HESS~J1846$-$029 \citep{dja08}.
However, Kes~75 differs from \ctb\ in that it 
contains a powerful X-ray PWN, while there is no evidence
of an X-ray PWN in \ctb.  Although these SNRs are of comparable
age, any X-ray PWN in \ctb\ may have evolved a factor of 10 faster
than the one in Kes~75, based on the scaling of pulsar $\dot E$
with true age $T$ such that $T \propto B_s^{-1}$ for a given $\dot E$.
The lifetime of a hypothetical rotation-powered X-ray PWN
around a magnetar must be relatively brief.  Interestingly,
\citet{vin10} claim to have detected the first X-ray PWN around 
a magnetar in an archival \chandra\ observation of
1E 1547.0$-$5408, the magnetar of highest $\dot E$.
Although this would be an important precedent, we hesitate
to interpret it as such because we have not
clearly confirmed its existence in those data.

Evidence to associate any other magnetar with TeV emission
is sparse.  The AXP 1E~1841$-$045 in SNR Kes~73 is located
at the edge of HESS~J1841$-$055 \citep{aha08b}, but this is not
a likely association because the TeV emission is $\approx 1^{\circ}$
in diameter and may conceivably have more than one source,
as considered by \citet{sgu09}. Moreover, it can be expected
that any high-energy particles are still contained within the SNR
shell, which is much smaller than the TeV source.
Recently, an extended TeV source that surrounds the massive,
young star cluster Westerlund~1 has also been
reported by HESS \citep{ohm09}.  It is possible that this source
is associated with a young pulsar born in the cluster, the transient
AXP CXO J164710.2$-$455216 \citep{mun06}, or its supernova remnant.
Notably absent among the reported HESS sources is
1E~1547.0$-$5408, the magnetar of highest $\dot E$
and a small characteristic age similar to \cco.
It would be interesting to target 1E~1547.0$-$5408
for a deep observation by HESS as a second test
of whether a young magnetar can produce a TeV source.

\section{X-ray Timing Observation of \psr\ in \tev}

In Paper~I, we also presented an analysis of the point source
\psr\ in the SNR \snr\ that has been identified with the TeV
source \tev\ \citep{aha08b,tia08}. We found only marginal evidence
for a 1~s period in an \xmm\ observation, which, in combination
with the absence of strong evidence for variability did not allow
us to definitively classify \psr.  Both a magnetar and a weakly
magnetized central compact object (CCO) remained possibilities.
The spectrum of \psr\ was best fitted by a two blackbody model,
which is common to members of both classes.

We obtained a new \chandra\ observation of \psr\ on 2010 May 18
using ACIS-S in CC-mode for 40~ks.  Using the same
analysis techniques as described in Section~2 for
CC-mode data, our period search of \psr\
did not reveal the candidate signal at 1~s, nor any other significant
period.  The 95\% confidence upper limit on the pulsed
fraction of a sinusoidal signal is 8.6\% for any period down to
10~ms.  If restricted to periods $\geq1$~s, the corresponding limit
is 7.6\%.  We also performed searches in restricted energy bands,
in case there is a pulsation with an energy-dependent phase reversal
similar to PSR J0821$-$4300 in Puppis~A \citep{got09},
but no such signal was found.

\begin{figure}[]
\centerline{\psfig{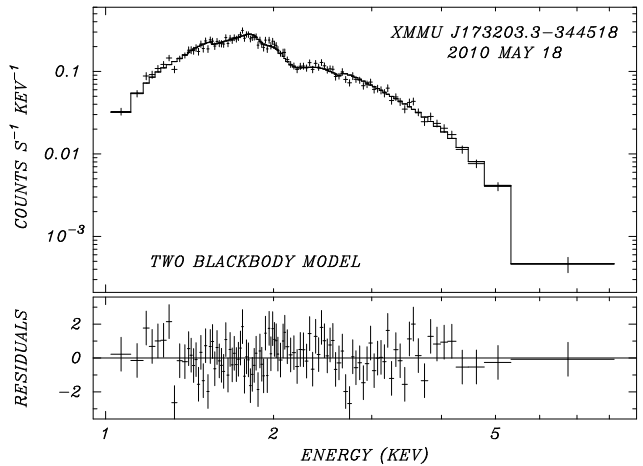}}
\caption{
Spectrum of \psr\ in \snr\ from the \chandra\ ACIS-S3 CC-mode
observation of 2010 May 18 fitted with two blackbodies.
Fitted parameters are listed in Table~\ref{spectable2}.
}
\label{spectrum2}
\end{figure}

We fitted the spectrum to a two blackbody model, which
was the best fitting model to previous observations
of \psr\ in Paper~I.
The result in shown in Figure~\ref{spectrum2}; the data is
grouped with 100 counts per bin.  Table~\ref{spectable2}
gives the parameters of the two blackbody fit.
These are consistent with the previous observations;
in particular, the 0.5$-$10~keV flux is identical to
that of an \xmm\ observation from the year 2007 (Paper~I).
The absence of variability and pulsations tends to
favor a CCO; several of the latter have very low limits
on pulsed fraction \citep[][and references therein]{hal10b}.
If so, \psr\ would be the most luminous CCO,
with $L_x \approx 1 \times 10^{34}$ erg~s$^{-1}$.
Recently, HESS was able to resolve
the TeV emission of \tev\ into a shell that matches the size
of the radio remnant \snr\ \citep{ace09}.
This would tend to rule out the
central source \psr\ as a source of the TeV emission.

\begin{deluxetable}{lc}
\tablewidth{0pt}
\tablecaption{\chandra\ Observation of \psr\ in \snr}
\tablehead{
\colhead{Parameter}  & \colhead{2010 May 18}
}
\startdata
ObsID                                     &  11234  \\
Mode                                      &  ACIS-S3/CC/F \\
Exp. time (s)                             & 39,920 \\
Counts (s$^{-1}$)                         &  0.334  \\
R.A. (J2000)\tablenotemark{a}         & $17^{\rm h}32^{\rm m}03^{\rm s}\!.40$ \\
Decl. (J2000)\tablenotemark{a}        & $-34^{\circ}45^{\prime}16^{\prime\prime}\!.77$ \\
\tableline
& Two Blackbody Model\\
\tableline
$N_{\rm H}$~($10^{22}$ cm$^{-2}$)         & $2.13^{+0.16}_{-0.14}$  \\
$kT_1$ (keV)                              & $0.36^{+0.04}_{-0.05}$  \\
$R_1$ (km)                                & $2.1_{-0.4}^{+1.3}$     \\
$kT_2$ (keV)                              & $0.61\pm 0.07$          \\
$R_2$ (km)                                & $0.44_{-0.29}^{+0.32}$  \\
$F_x(0.5-10\ {\rm keV}$)\tablenotemark{b} & $2.74 \times 10^{-12}$  \\
$L_x(0.5-10\ {\rm keV}$)\tablenotemark{c} & $1.26 \times 10^{34}$   \\
$L_{\rm bol}$\tablenotemark{b}            & $1.34 \times 10^{34}$   \\
$\chi^2_{\nu}(\nu)$                       & 1.066(102)               
\enddata
\tablenotetext{a}{Measured from \chandra\ ACIS-I ObsID 9139.
Typical \chandra\ ACIS coordinate uncertainty is $0\farcs6$.}
\tablenotetext{b}{Absorbed flux in units of erg cm$^{-2}$ s$^{-1}$.}
\tablenotetext{c}{Unabsorbed luminosity in units of erg s$^{-1}$,
assuming $d = 3.2$~kpc.}

\label{spectable2}
\end{deluxetable}

\section{Conclusions}
 
In Paper~I, we reported the discovery of
the 3.82~s pulsar \cco\ in \ctb,
which is coincident with the TeV source \tevtwo.
Additional X-ray timing measurements reported here
for the first time reveal the spin-down rate.  \cco\
is one of the most energetic magnetars
in terms of spin-down power, and has a characteristic
timing age of $\sim 1000$~years or less.
Its timing age is roughly consistent with
other estimates of the age of the SNR.
In addition, there is some
evidence for long-term variability by a factor of $\sim 2$
in spin-down rate, which recently is found at its highest
measured value.  Its $\dot E\/$ falls in the range
occupied by SGRs, which may portend a future SGR outburst.
The off-center location of the pulsar
inside the SNR shell is probably a result of
the inhomogeneous medium in which the shell expanded,
rather than indicating an unprecedented high natal
kick velocity.

There are considerable theoretical obstacles
to getting particles of TeV energy from a magnetar,
as well as having them emit primarily TeV $\gamma$-rays.
As well, there is little observational evidence for such particles
in the form of an X-ray synchrotron nebula.
Nevertheless, the exceptional youth and
high $\dot E$ of \cco\ lead us to entertain the
possibility that it generated a relic PWN
that is now emitting primarily via inverse Compton
scattering at TeV energies.  Future imaging
atmospheric Cherenkov observations to try to 
resolve the shell of \ctb\ from the pulsar
could test this hypothesis.

Turning to another compact source in a HESS SNR, 
we obtained a second X-ray timing observation
of \psr\ in \tev/\snr,
but did not succeed in discovering a period.
Its nature remains ambiguous because its composite
spectrum, best fitted with two blackbodies, is characteristic
of either AXPs or CCOs.  In comparison with three previous
X-ray observations of this source, the absence of obvious variability
so far is consistent with \psr\ being a CCO (anti-magnetar).
Deeper observations will be necessary to discover its timing
properties, which would determine conclusively
whether it is a magnetar or an anti-magnetar.
The present upper limit on pulsed
fraction is comparable to the weakest pulsations observed
in magnetars including, at times, 1E~1547.0$-$5408 \citep{hal08},
and similar to the upper limits on several CCOs \citep{hal10b}
for which pulsations have not been detected.  Even if
pulsations are absent, continuing monitoring of \psr\ 
for variability may also resolve its nature, since
only a magnetar would be variable.

\acknowledgements

Support for this work was provided by the National Aeronautics and Space
Administration through \chandra\ Awards GO9-0063X and GO0-11085X
issued by the \chandra\ X-ray Observatory Center, which is operated by the
Smithsonian Astrophysical Observatory for and on behalf of NASA under
contract NAS8-03060.

\end{document}